\documentclass[12pt]{article}
\usepackage{graphicx}
\usepackage{latexsym,amssymb}
\newcommand{\preprint}[1]{\begin{flushright}#1\end{flushright}}
\newcommand{\bea}{\begin{eqnarray}}
\newcommand{\eea}{\end{eqnarray}}
\newcommand{\simgt}{\hbox{ \raise3pt\hbox to 0pt{$>$}\raise-3pt\hbox{$\sim$} }}
\newcommand{\simlt}{\hbox{ \raise3pt\hbox to 0pt{$<$}\raise-3pt\hbox{$\sim$} }}

\begin{document}
\preprint{TU-594\\June 2000}
\vspace*{3cm}
\begin{center}
  {\bf\large
${\cal O}(\alpha_S^5 m)$
quarkonium $1S$ spectrum 
in large-$\beta_0$ approximation\\
and renormalon cancellation
}
  \\[10mm]
  {\large
    Y.~Kiyo and Y. Sumino
    }
  \\[10mm]
  {\it
    Department of Physics, Tohoku University\\
    Sendai, 980-8578 Japan
    }
\end{center}
\vspace{3cm}
\begin{abstract}
Presently the quarkonium spectrum,
written in terms of the quark $\overline{\rm MS}$-mass, is known
at ${\cal O}(\alpha_S^3 m)$ accuracy. 
We point out that in order to achieve ${\cal O}(\alpha_S^4 m)$ accuracy
it is sufficient to calculate further
(I) the ${\cal O}(\alpha_S^4 m)$ relation between the
$\overline{\rm MS}$-mass and the pole-mass, and
(II) the binding energy at ${\cal O}(\alpha_S^5 m)$ in the
large-$\beta_0$ approximation.
We calculate the latter correction analytically 
for the $1S$-state and study its
phenomenological implications.
\end{abstract}

\newpage
\section{Introduction}
\label{s1}

In recent years theory necessary for precise description
of a heavy quarkonium such as bottomonium or (remnant of)
toponium has developed significantly.
In particular, development in the computational technology of
higher order corrections to the quarkonium energy spectrum
\cite{py,my} and subsequent discovery of the 
renormalon cancellation \cite{renormalon1,renormalon2}
enabled accurate determinations
of the $\overline{\rm MS}$-mass of the bottom quark 
\cite{py,pp,my,upsilonmass} and
(in the future) of the top quark \cite{topcollab}.
In these determinations a major part is played by
the spectrum (mass) of the $1S$-state quarkonium.\footnote{
See e.g.\ \cite{pro,sumino} for introductory reviews of the subject.
}

It is legitimate to consider that the present perturbative calculation of
the quarkonium spectrum, when expressed in terms of the quark
$\overline{\rm MS}$-mass, has a 
{\it genuine accuracy} at ${\cal O}(\alpha_S^3 m)$.
In fact, in formal power countings, the last known term in
the relation between the
$\overline{\rm MS}$-mass and the pole-mass of a quark is 
${\cal O}(\alpha_S^3 m)$, while the last known term of
the binding energy (measured
from twice of the quark pole-mass) is
${\cal O}(\alpha_S^4 m)$.
The former term
includes in addition to a genuine ${\cal O}(\alpha_S^3 m)$
part the leading renormalon contribution which does not become
smaller than ${\cal O}(\Lambda_{\rm QCD})$ \cite{bbb}.
This renormalon contribution is cancelled \cite{renormalon1,renormalon2}
against the renormalon
contribution \cite{al} contained in the latter term.
Therefore, after cancellation of the leading renormalons,
the {\it genuine} ${\cal O}(\alpha_S^3 m)$ part of the mass relation
determines the accuracy of the present perturbation series
relating the quark $\overline{\rm MS}$-mass and the quarkonium
spectrum.

As stated, for the binding energy the calculation including the
{\it genuine} ${\cal O}(\alpha_S^4 m)$ corrections has already 
been completed.
Also, the 
``large-$\beta_0$ approximation''
\cite{bb}
is known to be a pragmatically feasible and empirically successful
estimation method of the leading renormalon contributions.
Taking these into account, one finds that it is sufficient\footnote{
There exist other methods in which the renormalon contribution contained
in the pole-mass is subtracted in certain approximations 
(e.g.\ \cite{renormalon2}).
In our opinion our method is a most natural one, embedded in the algorithm
for higher order calculations of the whole spectrum;
cancellation of infrared sensitivities in the whole spectrum follows
from the fact
that the quarkonium is a color-singlet small-size system.
} 
to calculate further the following two corrections
in order to improve the accuracy of
the spectrum by one order and to
achieve a genuine accuracy at ${\cal O}(\alpha_S^4 m)$:
(I) the ${\cal O}(\alpha_S^4 m)$ relation between the
$\overline{\rm MS}$-mass and the pole-mass, and
(II) the binding energy at ${\cal O}(\alpha_S^5 m)$ in the
large-$\beta_0$ approximation.
This is because the leading renormalon contribution in the full
${\cal O}(\alpha_S^5 m)$ correction to the binding energy will be
incorporated by the large-$\beta_0$ approximation and the
remaining part is expected to be irrelevant at ${\cal O}(\alpha_S^4 m)$.

Of these two corrections we calculate 
(II) analytically for the $1S$-state in this paper.
Then we study its phenomenological applications.
We can check validity of the above general argument
explicitly at ${\cal O}(\alpha_S^3 m)$ where we know the
exact result.
This will also be demonstrated.

Written in terms of the quark pole-mass $m_{\rm pole}$,
the mass of the quarkonium $1S$-state is given as a series expansion
in the $\overline{\rm MS}$ coupling constant $\alpha_S(\mu)$
defined in the theory with $n_l$ massless quarks:
\bea
M_{1S} = 2 m_{\rm pole}  - \frac{4}{9} \alpha_S(\mu)^2 m_{\rm pole}
\sum_{n=0}^\infty \Bigl( \frac{\alpha_S(\mu)}{\pi} \Bigr)^n
P_n (L) ,
\label{m1s}
\eea
where $P_n (L)$ is an $n$-th-degree polynomial of
$L \equiv \log [ 3 \mu/(4 \alpha_S(\mu) m_{\rm pole}) ]$.
At each order of the perturbative expansion
$c_n = P_n (0)$ represents a non-trivial correction, while
the coefficients of $L$'s are determined by the renormalization-group
equation.\footnote{
For $n \geq 3$,
$c_n$'s include powers of $\log \alpha_S$ unrelated to the
renormalization group, i.e.\ which is not accompanied
by $\log \mu$ \cite{kp}.
}
The polynomials relevant to our analysis read
\bea
P_0 &=& 1,
\\
P_1 &=& \beta_0  \, L + c_1 ,
\\
P_2 &=&
\frac{3}{4} \beta_0^2 \, {L^2} + 
  \left( - \frac{1}{2} \beta_0^2 
             + 
     {\frac{1}{4}\beta_1 } + 
     {\frac{3}{2} \beta_0 {c_1}} \right) L  + 
  {c_2} ,
\\
P_3 &=&
\frac{1}{2}\beta_0^3  \, L^3 + 
  \left( -\frac{7}{8}\beta_0^3
        + \frac{7}{16}\beta_0\beta_1
          + 
     \frac{3}{2}\beta_0^2c_1 \right) L^2
\nonumber \\ 
&& ~~~
      + \left( \frac{1}{4}\beta_0^3
        - \frac{1}{4}\beta_0\beta_1
          + 
     \frac{1}{16}\beta_2 - 
     \frac{3}{4}\beta_0^2c_1 + 
     \frac{3}{8}\beta_1c_1 + 
     2\beta_0c_2 \right)L  + c_3 .
\eea
$\beta_n$'s denote the coefficients of the QCD beta function
given by
\bea
&&
\beta_0 = 11 - \frac{2}{3} n_l ,
~~~~~~~~~~
\beta_1 = 102 - \frac{38}{3} n_l,
~~~~~~~~~~
\beta_2 = \frac{2857}{2} - \frac{5033}{18}n_l
+ \frac{325}{54} n_l^2 .
\eea
Note that, in the renormalization-group evolution, 
running of 
the coupling included in $\log \alpha_S$
should be taken into account properly.
This feature is unique to the perturbation series of
a nonrelativistic boundstate spectrum.
Presently $c_n$'s are known up to $n=2$ \cite{py,my}:
\bea
c_1 &=& \frac{97}{6}-\frac{11}{9}n_l ,
\label{knownres1}
\\
c_2 &=& 
\frac{1793}{12} + \frac{2917\pi^2}{216}
-\frac{9\pi^4}{32}+\frac{275 \zeta_3}{4} 
+ \Bigl( - \frac{1693}{72}-\frac{11\pi^2}{18}-\frac{19\zeta_3}{2} 
\Bigr)n_l
+ \Bigl( \frac{77}{108}+\frac{\pi^2}{54}+\frac{2\zeta_3}{9} \Bigr)
n_l^2 .
\nonumber \\
\label{knownres2}
\eea
The aim of this paper is to calculate $c_3$ in the large-$\beta_0$
approximation.

One might think that alternatively $c_3$ may be
estimated
using the asymptotic form of the series expansion of the QCD potential
at large orders.
We could not, however, find a justification that $c_3$ can be
estimated with ${\cal O}(\alpha_S^4 m)$ accuracy in this
manner.

\section{Outline of the calculation}
\label{s2}

In the large-$\beta_0$ approximation the wave functions
and energy spectra (measured from $2m_{\rm pole}$) of quarkonium
states
are determined by solving the nonrelativistic Schr\"odinger equation
\bea
\left[
\frac{\vec{p}\, ^2}{m} + V_{\beta_0}(r)
\right] \psi (\vec{x}) = E \psi (\vec{x}) .
\eea
Here, $m$ denotes the mass of the quark and it is irrelevant
whether we use the pole-mass or the $\overline{\rm MS}$-mass
for $m$ within our approximation;
in particular it does not affect the leading renormalon cancellation.
$ V_{\beta_0}$ denotes the QCD potential in the large-$\beta_0$
approximation given as a perturbation series \cite{al}
\bea
V_{\beta_0}(r) = - C_F \, \frac{\alpha_S(\mu)}{r} \times R(r) ,
~~~~~~~~~
R(r) = \sum_{n=0}^\infty 
\Biggl( \frac{\beta_0 \alpha_S(\mu)}{4\pi} \Biggr)^n
R_n(\mu r) .
\label{R}
\eea
$C_F = 4/3$ is the color factor.
First few coefficients $R_n(\mu r)$ are listed in
Table~\ref{t1}.
\begin{table}
\begin{center}
\begin{tabular}{c|c}
\hline
$n$ & $R_n(\mu r)$ \\
\hline
 0  & $1$     \\
 1  & $2 \ell $  \\
 2  & $4 \ell^2 + \frac{\pi^2}{3} $  \\
 3  & $8 \ell^3 + 2\pi^2 \ell + 16 \zeta_3 $  \\
 $\vdots$ & $\vdots$ \\
\hline
\end{tabular}
\caption{\small
First few coefficients of the perturbative
expansion of $R(r)$.
Here, $\ell = \log (\mu r) + \frac{5}{6} + \gamma_E$.
}
\label{t1}
\end{center}
\end{table}
To obtain $S$-wave solutions, 
we introduce a dimensionless variable
$z = C_F \alpha_S m r$ and set
\bea
\psi  = \frac{1}{z} \, \exp
\left[ - \int dz \, W(z) \right],
~~~~~~~~~
E = (C_F \alpha_S)^2 m \varepsilon .
\eea
Then the equation becomes
\bea
W' - W^2 - R/z = \varepsilon .
\label{weq}
\eea
We expand $W$, $R$ and $\varepsilon$ in perturbation series
as in (\ref{R}) and
\bea
W = \sum_{n=0}^\infty 
\Biggl( \frac{\beta_0 \alpha_S}{4\pi} \Biggr)^n
W_n ,
~~~~~~~~~~
\varepsilon = \sum_{n=0}^\infty 
\Biggl( \frac{\beta_0 \alpha_S}{4\pi} \Biggr)^n
\varepsilon_n ,
\eea
and substitute them to (\ref{weq}).
We find
\bea
W_n' - \sum_{k=0}^{n} W_k \, W_{n-k} - R_n/z = 
\varepsilon_n .
\label{wneq}
\eea
For the $1S$-state the zeroth-order solution is given by
\bea
W_0 = \frac{1}{2} - \frac{1}{z},
~~~~~~~~~
\varepsilon_0 = - \frac{1}{4} ,
\eea
and we may solve (\ref{wneq}) recursively for $n \ge 1$:
\bea
\varepsilon_n &=& - \frac{1}{2}
\int_0^\infty dt \,
t^2 e^{-t}
\Biggl(
\sum_{k=1}^{n-1} W_k(t) \, W_{n-k}(t) + R_n(t)/t
\Biggr) ,
\\
W_n(z) &=& - \frac{e^z}{z^2} \int_z^\infty dt \,
t^2 e^{-t}
\Biggl(
\sum_{k=1}^{n-1} W_k(t) \, W_{n-k}(t) + R_n(t)/t + \varepsilon_n
\Biggr) .
\eea
Thus, $\varepsilon_n$'s can be obtained by evaluating
multiple integrals.
We may render these integrals 
to forms which resemble Feynman-parameter integrals that appear
in calculations of multiloop Feynman diagrams.
In this way we could use various techniques developed for Feynman
diagram calculations.\footnote{
We find the techniques developed in \cite{broadhurst} particularly useful.
}
We obtained the first two terms of the perturbative expansion as
\bea
\varepsilon_1 &=& - ( \tilde{L} + 1 ) ,
\\
\varepsilon_2 &=& - ( 3 \tilde{L}^2 + 4 \tilde{L} + 1 +
                      \frac{\pi^2}{6} + 2 \zeta_3 ) ,
\eea
where $\tilde{L} = L + \frac{5}{6}
= \log [ \mu/(C_F \alpha_S m) ] + \frac{5}{6}$.
From these we confirmed the corresponding parts of $c_1$ and
$c_2$ in eqs.~(\ref{knownres1},\ref{knownres2}).
We also obtained a new result:
\bea
\varepsilon_3 = - \Biggl[
8  {\tilde{L}^3} + 10  {\tilde{L}^2}  + 
  \left( {\frac{4  {{\pi }^2}}{3}} + 
     16  \zeta_3 \right)  \tilde{L}  
-2  + {{\pi }^2} + 16  \zeta_3 +
  {\frac{{{\pi }^4}}{90}} - 
  2  {{\pi }^2}  \zeta_3 + 24  \zeta_5
\Biggr] .
\eea
It follows that
\bea
c_3(\mbox{large-}\beta_0) = \beta_0^3 \, \Biggl(
{\frac{517}{864}} + {\frac{19\,{{\pi }^2}}{144}} + 
  {\frac{11\,{\zeta_3}}{6}} + 
  {\frac{{{\pi }^4}}{1440}} - 
  {\frac{{{\pi }^2}\,{\zeta_3}}{8}} + 
  {\frac{3\,{\zeta_5}}{2}}
\Biggl) .
\label{c3beta0}
\eea
We note two points:
(i) The result includes the level-5 zeta values
($\zeta_5$ and $\pi^2 \zeta_3$).
(ii) The result does not include $\gamma_E$, that is,
all $\gamma_E$ which appeared at intermediate stages of
the calculation got cancelled.
A more detailed description of our calculation will be published
elsewhere.

\section{Phenomenological applications}

We examine the series expansion of the quarkonium 
$1S$-state spectrum
numerically for the bottomonium and (remnant of)
toponium.
The input value for the coupling defined in the theory
with 5 massless flavors ($n_l=5$)
is $\alpha_S^{(5)}(M_Z) = 0.119$.
We evolve the coupling and match it to
the coupling of the theory with $n_l=4$ 
following \cite{running}.
Also we take the input values of the $\overline{\rm MS}$-mass,
$\overline{m} \equiv m_{\overline{\rm MS}}(m_{\overline{\rm MS}})$,
and the pole-mass as
$\overline{m}_b = 4.20$~GeV/$m_{b,{\rm pole}} = 4.97$~GeV and
$\overline{m}_t = 165$~GeV/$m_{t,{\rm pole}} = 174.79$~GeV.
The natural size of a quarkonium system is the Bohr radius.
We define a corresponding scale parameter $\mu_B$
such that
\bea
\mu_B = C_F \alpha_S(\mu_B) m_{\rm pole}
\eea
holds.
We examine the series expansion (\ref{m1s}) with two different choices of 
the scale $\mu$ in Table~\ref{tab2}.
\begin{table}[t]
\begin{center}
\small
\begin{tabular}{c|r|r}
\hline
\multicolumn{3}{c}{
Series expansion of $M_{1S}$ [GeV]
} \\
\hline \hline
& \multicolumn{1}{c|}{$\mu = \mu_B$} & 
\multicolumn{1}{c}{$\mu = \overline{m}$}
\\
& \multicolumn{1}{c|}{[~expansion parameter $= \alpha_S(\mu_B)$~]} 
& \multicolumn{1}{c}{[~expansion parameter $= \alpha_S(\overline{m})$~]} 
\\
\hline
Bottomonium & 
$2 \times (
4.97 -  0.11  -  0.12  -  0.20  -  0.25^\star  
)$ &
$2 \times (
4.97 -  0.06  -  0.08  -  0.11  -  0.15^\star  
)$ 
\\
$(n_l=4)$ & $[ - \, 0.26 ]~\,$ & $[- \, 0.16]$~\,
\\
\hline
Toponium & 
$2 \times (
174.79 -  0.77  -  0.35  -  0.25  -  0.13^\star  
)$ &
$2 \times (
174.79 -  0.46  -  0.39  -  0.28  -  0.19^\star 
)$ 
\\
$(n_l=5)$ & $[ - \, 0.14 ]~\,$ & $[- \, 0.25]$~\,
\\
\hline
\end{tabular}
\caption{\small
Numerical evaluation of the series expansion of $M_{1S}$,
eq.~(\ref{m1s}).
Relevant parameters for the bottomonium are:
$\overline{m}_b = 4.20$~GeV, $m_{b,{\rm pole}} = 4.97$~GeV,
$\mu_B=2.05$~GeV, $\alpha_S^{(4)}(\mu_B)=0.309$,
$\alpha_S^{(4)}(\overline{m}_b)=0.229$;
those for the toponium are:
$\overline{m}_t = 165$~GeV, $m_{t,{\rm pole}} = 174.79$~GeV,
$\mu_B=32.9$~GeV, $\alpha_S^{(5)}(\mu_B)=0.1411$,
$\alpha_S^{(5)}(\overline{m}_t)=0.1092$.
}
\label{tab2}
\end{center}
\end{table}
The last terms (with stars) are evaluated using the value of
$c_3$ in the large-$\beta_0$ approximation (\ref{c3beta0}).
One sees that for the bottomonium 
the series expansions do not converge at all.
For the toponium the series expansions converge very slowly.
The numbers in square brackets represent
estimates of the last terms using
the asymptotic form of the series expansion of 
the potential $V_{\beta_0}(r)$ \cite{al}:
\bea
%{\textstyle
- \, C_F \, \frac{\alpha_S(\mu)}{r} \times
%\Bigl(  \frac{\beta_0 \alpha_S(\mu)}{4\pi} \Bigr)^n
R_n(\mu r) \sim
- \, \frac{2 \, e^{5/6} C_F \alpha_S(\mu) \, \mu}{\pi}
\times 2^n \, n! 
%\Bigl( \frac{\beta_0 \alpha_S(\mu)}{2\pi} \Bigr)^n
~~~~~
\mbox{for}~
n \gg 1 .
%\nonumber \\
\label{asymptoticf}
%}
\eea
The series approaches its asymptotic form 
faster when
we choose $\mu=\mu_B$.
These features are consistent with 
dominance of the leading renormalon contributions.

Next we rewrite the series expansion of $M_{1S}$ in
terms of the $\overline{\rm MS}$-mass instead of the
pole-mass.
The leading renormalon contributions cancel in this case.
Presently the relation between the $\overline{\rm MS}$-mass and
the pole-mass is known up to three loops \cite{polemass1,polemass2}.
The only scale in this relation is the quark mass.
Thus, we
have two choices of scales, $\overline{m}$ and $\mu_B$, 
in writing the series expansion of $M_{1S}$;
we take $\mu = \overline{m}$ below.
We eliminate the pole-mass completely and expand
in $\alpha_S(\overline{m})$.
We should properly take into account the fact that
the renormalon contributions cancel between
the terms whose orders in $\alpha_S$ differ by one \cite{hlm}.
To this end we proceed as follows.
We rewrite (\ref{m1s}) as
\bea
M_{1S} &=& 2 m_{\rm pole} \times
\biggl[ \, 1 + \sum_{n=2}^\infty \, Q_n \, \alpha_S(\overline{m})^n
\biggr]
\nonumber \\
&=&
2 \overline{m} \times 
\biggl[ \, 1 + \sum_{n=1}^\infty \, d_n \, \alpha_S(\overline{m})^n
\biggr] \times
\biggl[ \, 1 + \sum_{n=2}^\infty \, Q_n \, \alpha_S(\overline{m})^n
\biggr] ,
\eea
where $Q_n$'s are polynomials of 
$\log [ \alpha_S(\overline{m}) ]$ and
$d_n$'s (the $n$-loop coefficients of the mass relation)
are just constants independent of $\alpha_S(\overline{m})$.
We identified
$Q_n \alpha_S^n$ as order $\alpha_S^{n-1}$
and then reduced the last line to a single series in $\alpha_S$.
Numerically we find\footnote{
The values of $d_1 \sim d_3$ are taken from eq.~(14) of
\cite{polemass2}.
}
\bea
M_{1S} &=&
2 \times ( 4.20 + 0.36 + 0.13 + 0.040 + 0.0051^\sharp )~{\rm GeV}
~~~~~~~
({\rm Bottomonium}) ,
\label{improvedser1}
\\
M_{1S} &=&
2 \times ( 165.00 + 7.21 + 1.24 + 0.22 + 0.052^\sharp )~{\rm GeV}
~~~~~~~
({\rm Toponium}) .
\label{improvedser2}
\eea
The last terms (with sharps) are evaluated using the values of
$c_3$ and $d_4$ \cite{bb} in the large-$\beta_0$ approximation.
Convergences of the series improve markedly in comparison to those
in Table~\ref{tab2}.
(Previous analyses similar to the one presented above can be found
in \cite{pro,topcollab} and references therein.)

As we argued in Section~\ref{s1}, parametric accuracy of the last
terms in (\ref{improvedser1},\ref{improvedser2}) 
is ${\cal O}(\alpha_S^4 m)$ and we need to know further only
the exact value of $d_4$ to make a perturbative evaluation accurate
up to this order (the exact form of $c_3$ is not necessary).
In order to verify validity of this argument, we replace $c_2$ by its
value in the large-$\beta_0$ approximation.
Then the ${\cal O}(\alpha_S^3 m)$ terms of 
(\ref{improvedser1},\ref{improvedser2})
change to 0.043 and 0.22, respectively.
Thus, we do not lose accuracy at this order by the replacement.
On the other hand, if we replace $c_2$ and $d_3$ by their
values in the large-$\beta_0$ approximation, the same
terms change to 0.056 and 0.31, respectively.
Thus, we lose the accuracy at ${\cal O}(\alpha_S^3 m)$.
These aspects are consistent with our general argument.
Also, they suggest that the last terms
of (\ref{improvedser1},\ref{improvedser2})
would be reasonable estimates of the orders of magnitude of
the exact ${\cal O}(\alpha_S^4 m)$ terms.

Finally we examine if we can use
the asymptotic form of $V_{\beta_0}(r)$, eq.~(\ref{asymptoticf}),
to estimate $c_3(\mbox{large-}\beta_0)$.
From the asymptotic values of the last terms 
for $\mu = \mu_B$ in Table~\ref{tab2},
we may extract approximate values of $c_3(\mbox{large-}\beta_0)$ as
\bea
c_3({\rm asympt}) = \left\{
\begin{array}{cc}
2.55\times 10^3 & (n_l=4) \\
1.98\times 10^3 & (n_l=5)
\end{array}
\right. ,
\eea
while the corresponding values of $c_3(\mbox{large-}\beta_0)$
are, respectively,  $2.46\times 10^3$ and  $1.91\times 10^3$.
If we substitute $c_3({\rm asympt})$, the last terms of 
(\ref{improvedser1},\ref{improvedser2})
change to 0.0034 and 0.051, respectively.
Therefore, we see that in the case of bottomonium
the use of the asymptotic form does not reproduce our result
in the large-$\beta_0$ approximation (\ref{improvedser1})
with good (relative) accuracy.
Presumably it is a sign of the next-to-leading
renormalon contribution, which does not become smaller than
$\sim \Lambda_{\rm QCD} \cdot ( \Lambda_{\rm QCD} / \mu_B )^2
\sim 2$~MeV, and which is not included in $c_3({\rm asympt})$.
Namely, we conjecture that already
the last term in eq.~(\ref{improvedser1})
stands close to the limit where an improvement of convergence of
the perturbation series is possible by cancellation of 
the leading renormalon contributions.

\section{Conclusions and discussions}

We have calculated the ${\cal O}(\alpha_S^5 m)$ correction
to the quarkonium $1S$-state energy spectrum
analytically in the large-$\beta_0$
approximation.
As a result, in order to predict the $1S$-state
spectrum at ${\cal O}(\alpha_S^4 m)$ perturbative
accuracy, only the four-loop relation between the
$\overline{\rm MS}$-mass and the pole-mass remains to be
computed.
Within the present approximation, the perturbation series of the
bottomonium and toponium spectra show healthy
convergent behaviors up to a genuine ${\cal O}(\alpha_S^4 m)$ accuracy.

In the case of bottomonium, current theoretical 
uncertainty due to non-perturbative effects is estimated
to be of the order of 0.1~GeV \cite{upsilonmass,pro}.
It is much larger than the size of the
last term of (\ref{improvedser1}).
We hope that in the future the theoretical uncertainty due to
non-perturbative effects will be reduced by applications
of e.g.\ lattice calculations or combinations of
operator-product-expansion and sum rules.

Part of the {\it genuine} ${\cal O}(\alpha_S^5 m)$
corrections have already been calculated \cite{kp}.
Taking $\mu = \overline{m}$, their individual sizes are
evaluated to be
$\Delta M_{1S} \sim \pm 0.05$~GeV
both for the bottomonium and toponium;
if we take $\mu = \mu_B$, they
become even an order of magnitude larger.
We do not know as yet whether it may indicate a breakdown of
perturbative expansion of the spectrum
at ${\cal O}(\alpha_S^5 m)$ or a requisiteness for
some new cancellation mechanism.
It is also possible that the sum of all the 
genuine ${\cal O}(\alpha_S^5 m)$
corrections turns out to be much smaller.
In any case, it would be better to separate the discussion
of this problem from the determination of the genuine
${\cal O}(\alpha_S^4 m)$ corrections
(as we advocated in this paper),
rather than to regard them as inseparable constituents of $c_3$.

\section*{Acknowledgements}

Y.K. was supported 
by the Japan Society for the Promotion of Science.
Y.S. was supported in part
by the Japan-German Cooperative Science
Promotion Program.

\def\app#1#2#3{{\it Acta~Phys.~Polonica~}{\bf B #1} (#2) #3}
\def\apa#1#2#3{{\it Acta Physica Austriaca~}{\bf#1} (#2) #3}
\def\npb#1#2#3{{\it Nucl.~Phys.~}{\bf B #1} (#2) #3}
\def\plb#1#2#3{{\it Phys.~Lett.~}{\bf B #1} (#2) #3}
\def\prd#1#2#3{{\it Phys.~Rev.~}{\bf D #1} (#2) #3}
\def\pR#1#2#3{{\it Phys.~Rev.~}{\bf #1} (#2) #3}
\def\prl#1#2#3{{\it Phys.~Rev.~Lett.~}{\bf #1} (#2) #3}
\def\sovnp#1#2#3{{\it Sov.~J.~Nucl.~Phys.~}{\bf #1} (#2) #3}
\def\yadfiz#1#2#3{{\it Yad.~Fiz.~}{\bf #1} (#2) #3}
\def\jetp#1#2#3{{\it JETP~Lett.~}{\bf #1} (#2) #3}
\def\zpc#1#2#3{{\it Z.~Phys.~}{\bf C #1} (#2) #3}

\end{document}